\documentstyle[aps,epsfig,float,amstex,axodraw]{revtex}

\newcommand{\be}{\begin{equation}}
\newcommand{\ee}{\end{equation}}
\newcommand{\beq}{\begin{eqnarray}}
\newcommand{\eeq}{\end{eqnarray}}

\newcommand{\al}{\ArrowLine}
\newcommand{\dl}{\DashLine}
\newcommand{\ve}{\Vertex}

\restylefloat{figure}
\restylefloat{table}

\begin{document}
    
\def\gC{\mbox{\boldmath $C$}}
\def\gZ{\mbox{\boldmath $Z$}}
\def\gR{\mbox{\boldmath $R$}}
\def\gN{\mbox{\boldmath $N$}}
\def\ua{\uparrow}
\def\da{\downarrow}
\def\eq{\equiv}
\def\a{\alpha}
\def\b{\beta}
\def\g{\gamma}
\def\G{\Gamma}
\def\d{\delta}
\def\D{\Delta}
\def\e{\epsilon}
\def\z{\zeta}
\def\h{\eta}
\def\th{\theta}
\def\k{\kappa}
\def\l{\lambda}
\def\L{\Lambda}
\def\m{\mu}
\def\n{\nu}
\def\x{\xi}
\def\X{\Xi}
\def\p{\pi}
\def\P{\Pi}
\def\r{\rho}
\def\s{\sigma}
\def\S{\Sigma}
\def\t{\tau}
\def\f{\phi}
\def\vf{\varphi}
\def\F{\Phi}
\def\c{\chi}
\def\w{\omega}
\def\W{\Omega}
\def\Q{\Psi}
\def\q{\psi}
\def\de{\partial}
\def\inf{\infty}
\def\ra{\rightarrow}
\def\bra{\langle}
\def\ket{\rangle}

\title{Pairing Mechanism in the doped Hubbard Antiferromagnet:  
the $4 \times 4$ Model as a Test Case}

\author{Michele Cini, Enrico Perfetto and Gianluca Stefanucci}

\address{Istituto Nazionale di Fisica della Materia, Dipartimento di Fisica,\\
Universita' di Roma Tor Vergata, Via della Ricerca Scientifica, 1-00133\\
Roma, Italy}
\maketitle

\begin{abstract}     
We introduce a {\em local} formalism, in terms of eigenstates of number 
operators, having well defined point symmetry,  to solve the Hubbard  model 
at weak coupling on a  $N\times N$ square lattice  (for even $N$).
The key concept is that of $W=0$ states, that are the  many-body 
eigenstates of the kinetic energy  with 
vanishing  Hubbard repulsion.  At half filling, the wave function 
demonstrates an antiferromagnetic order, a lattice step translation 
being equivalent to a spin flip.  Further,
we state a general theorem which allows to find all the  
$W=0$ pairs (two-body $W=0$ singlet states). 
We show that, in special cases, this assigns the ground 
state symmetries at least in the weak coupling regime. 
The $N=4$ case is discussed in detail. To study the doped half filled system, 
we enhance the Group Theory analysis of the 
$4 \times 4$ Hubbard model introducing an Optimal Group which 
explains all the degeneracies in the one-body and many-body spectra. 
We use the Optimal Group to  predict the possible ground state symmetries 
of the  $4 \times 4$ doped antiferromagnet by means of our general theorem 
and the results are in agreement with exact diagonalization data.
Then we create $W=0$ {\it electron} pairs over the 
antiferromagnetic state. We show analitycally that the effective interaction 
between the electrons of the pairs is attractive and forms  
bound states. Computing the corresponding binding energy we are able to definitely 
predict the exact ground state symmetry.
\end{abstract}

\section{Introduction}
\label{intro}

The  (repulsive) $2D$  Hubbard hamiltonian   is one of the popular models of the 
high-$T_{C}$ cuprates\cite{bm}, as many people believe that it contains 
at least some of the relevant ingredients of the mechanism of superconductivity. 
While other ingredients may well be missing for the full explanation of 
superconductivity and the rich phase diagram of these materials, 
there are now strong evidences of pairing in this model, and 
although pairing is not synonimous to superconductivity it can 
hardly be supposed to be extraneous to it. The evidence 
for pairing comes from various independent methods, including cluster 
diagonalizations\cite{fop}\cite{cibal}\cite{fm},
fluctuation exchange  (FLEX)\cite{ms} diagrammatic approach, which is based on a conserving
approximation, and renormalization group 
techniques\cite{zs}\cite{hm}.  The approach we are 
proposing is based on an analytic canonical transformation, 
reminiscent of the original Cooper theory. However, our understanding 
of the many-body ground state is only partial, and the symmetries of 
the  bound pairs have not yet been fully explored.  The Hubbard model must be more thoroughly 
understood before we can solve more realistic ones. 

In the strong coupling limit the double occupation of the same 
site is energetically suppressed and the model at half filling is 
equivalent to the Heisenberg model with an antiferromagnetic exchange 
interaction\cite{a}. A popular approach takes care of the strong repulsion 
between two opposite spin fermions by a Gutzwiller\cite{Gutzwiller} projection, 
i.e. by  throwing out of the Hilbert space  the  double occupation states.
However, truncating the Hilbert space in this way costs kinetic 
energy, so  at finite $U$ the system must allow double occupation, 
also in  the ground state, as one can see from the eigenvectors of 
cluster calculations.  At weak coupling, on the other hand, it makes sense to speak 
about particles in filled shells, which behave much as core electrons 
in atomic physics, and particles in partially filled, or valence, shells.  
Remarkably, particles in partially filled  shells can \underline{totally} avoid double occupation at \underline {no cost} in energy; they 
do so, forming {\bf W=0 states,} that are defined as {\bf many-particle eigenstates of the 
kinetic energy with no double occupation.} Below, using a new 
formalism, we show  how $W=0$ states arise by symmetry.  We stress 
that since $W=0$ states  emerge from  symmetry alone, they remain 
$W=0$ for any coupling strength and are an adequate starting point 
for a realistic theory. This is the reason why  weak coupling 
expansions often provide good approximations at intermediate 
coupling, as observed by several Authors \cite{metzner}\cite{fridman}\cite{galan}.   

\subsection{Canonical Transformation Approach to the Pairing Mechanism}
\label{paimec}

Let us consider the Hubbard model with hamiltonian 
\begin{equation}
H=H_{0}+W=t\sum_{\s}\sum_{\bra r,r'\ket}c^{\dag}_{r\s}c_{r'\s}+
\sum_{r}U\hat{n}_{r\ua}\hat{n}_{r\da},\;\;\;\;U>0,
\label{hamil}
\end{equation}
on a square lattice of $N\times N$ sites with periodic boundary 
conditions and even $N$. Here $\s=\ua,\da$ is the 
spin and $r,\;r'$ the spatial  degrees of freedom of the hole creation and 
annihilation operators $c^{\dag}$ and $c$ respectively. The sum on 
$\bra r,r'\ket$ is over the pairs of nearest neighbors sites and 
$\hat{n}_{r\s}$  is the number operator on the site 
$r$ of spin $\s$. The point symmetry is 
$C_{4v}$, the Group of a square\cite{appendix}; besides, $H$ is 
invariant under the  commutative Group of Translations ${\mathbf 
T}$ and hence the Space Group\cite{hamer}   ${\mathbf G}={\mathbf 
T} \otimes C_{4v} $; $\otimes$ means the semidirect product. We   
represent sites by $r=(i_{x},i_{y})$ and wave vectors by
$k=(k_{x},k_{y})=\frac{2\p}{N}(i_{x},i_{y})$,
with  $i_{x},i_{y}=0,\ldots,N-1$. In terms of the 
Fourier expanded fermion operators 
$c_{k\s}=\frac{1}{N}\sum_{r}e^{ikr}c_{r\s}$, we have 
$H_{0}=\sum_{k}\e(k)c^{\dag}_{k\s}c_{k\s}$ with $\e(k)=2t
[\cos k_{x}+\cos k_{y}]$. Then the one-body plane wave state $c^{\dag}_{k\s}|0\ket\equiv|k\s\ket$ 
is an  eigenstate of $H_{0}$.

To study the behaviour of two holes added to the system 
in its ground state, we introduce the  $W=0$ pairs, a special case of $W=0$ states. 
Using degenerate eigenstates of the kinetic energy 
$H_{0}$, the non-abelian point symmetry 
Group $C_{4v}$ of the Hubbard hamiltonian (\ref{hamil}) allows the 
existence of two-body singlet states with no double occupancy. They 
are obtained by a configuration interaction mechanism and 
may be of special interest since the particles of a $W=0$ pair do not 
feel any direct repulsion. Hence a new pairing scenario  arises: since the two 
extra particles cannot interact directly, by definition of $W=0$ pair, 
their effective interaction comes out from virtual electron-hole 
excitations exchanges with the {\it Fermi sea} and in priciple can be 
attractive. In the following we show how to get the effective 
interaction between two holes added to the system.

Many configurations contribute to 
the  interacting $(n+2)$-body ground state $|\Psi _{0}(n+2)\ket$ and we need a complete 
set $\mathcal{S}$ 
to expand it exactly; as long as it is complete, however, we can 
design $\mathcal{S}$  as we please.  We can take the non-interacting 
$n$-body Fermi {\em sphere}
$|\Phi(n)\ket$ as our vacuum and build the complete set  in terms 
of excitations over the vacuum. 
In the subspace with vanishing spin $z$ component, 
the simplest states  that enter the configuration mixing are those  obtained from  
$|\Phi(n)\ket$ by creating two extra holes over it (we call them 
the $m$ states); the set
\begin{equation}
c^{\dag}_{k_{1}\ua}c^{\dag}_{k_{2}\da}|\F(n)\ket,\;\;\;\;\e(k_{1}),\e(k_{2})>\e_{F}
 \label{statim}
\end{equation} 
where $\e_{F}$ is the Fermi energy, is a basis for this part of the Hilbert space. 
Similarly, along with the  pair $m$ states, we introduce the  4-body $\alpha $  
states, obtained from $|\Phi(n)\ket$ by creating $2$ holes and 1 
electron-hole (e-h) pair;  a basis looks like
\begin{equation}
c^{\dag}_{k_{1}\ua}c^{\dag}_{k_{2}\da}c_{k_{3}\s}c^{\dag}_{k_{4}\s}|\F(n)\ket,
\;\;\;\;\;\e(k_{1}),\e(k_{2}),\e(k_{4})>\e_{F}>\e(k_{3}),\;\;\s=\ua,\da
\label{statia}
\end{equation}
Then  $\mathcal{S}$  includes 
the  6-body $\beta $  states  having $2$ holes and 2 e-h pairs, and so 
on until states with two holes and $n$ e-h pairs if we are below half 
filling or $2N^{2}-n$ e-h pairs if we are above. 
We are using Greek indices for the configurations 
containing the electron-hole pairs, which here are playing largely the 
same r$\hat{o}$le as phonons in the Cooper theory. 
By means of the complet set $\mathcal{S}$ we now expand the 
interacting ground state 
\begin{equation}
|\Psi _{0}(n+2)\ket={\sum_{m}}a_{m}|m\ket+{\sum_{\alpha }}b_{\alpha }|\alpha \ket+{\
\sum_{\beta }}c_{\beta }|\beta \ket+....  
\label{lungo}
\end{equation}
and set up the Schr\"{o}dinger equation
\begin{equation}
H|\Psi _{0}(n+2)\ket=E|\Psi _{0}(n+2)\ket.
\label{seq}
\end{equation}
We stress that Eq.(\ref{lungo}) is configuration interaction, 
{\em not a perturbative expansion}. 

When  the number $n$  of holes in the 
$N\times N$ system is such that $|\Phi(n)\ket$ is a single 
non-degenerate determinant (the Fermi surface is totally filled),  we 
can easily and unambiguously define and calculate  the effective interaction between the 
two extra holes since the expansion (\ref{lungo}) for the interacting 
ground state is unique: this is done  by a canonical 
transformation\cite{cbs1},\cite{cbs2},\cite{cbs3} from the 
many-body Hamiltonian of Eq.(\ref{hamil}). We consider the effects of the
operators on the terms of $|\Psi _{0}(n+2)\ket$. We write:
\begin{equation}
H_{0}|m\ket=E_{m}|m\ket,  
\label{h0effect}
\end{equation}
and since $W$ can create or destroy up to 2 e-h pairs, its action on 
an $m$ state yields
\begin{equation}
W|m\ket={\sum_{m^{\prime }}}W_{m^{\prime },m}|m^{\prime }\ket+{\sum_{\alpha }}
|\alpha\ket W_{\alpha ,m} 
+{\ \sum_{\beta }}|\beta\ket W_{\beta ,m}.  
\label{weffect}
\end{equation}
The matrix elements of the two-body interaction 
$W$ between determinants which differ by two spin-orbitals 
reduces to a bielectronic integral. Hence one can show that  if
$m^{\prime },m$ represent different $W=0$ pairs added to the vacuum, 
$W_{m^{\prime },m}$ vanishes. Moreover,
\begin{equation}
W|\alpha\ket={\sum_{m}}|m\ket W_{m,\alpha }+{\sum_{\alpha ^{\prime }}}|\alpha
^{\prime }\ket W_{\alpha ^{\prime },\alpha }  
+{\sum_{\beta }}|\beta\ket W_{\beta ,\alpha } +\sum_{\g}|\g\ket W_{\g\a},
\label{weffect2}
\end{equation}
where scattering between 4-body states is allowed by the second term, 
and so on. In this way we obtain an algebric system for the 
coefficients of the configuration interaction (\ref{lungo}). However to test 
the instability of the Fermi liquid towards pairing it is 
sufficient to study the amplitudes $a_{m}$ of the $m$ states. In the 
weak coupling limit this can be done truncating the expansion (\ref{lungo}) 
to the $\a$ states because, as we have shown\cite{cbs1}, the inclusion 
of the $\b,\g,\ldots$ states 
produces a renormalization of the matrix elements of higher order in $W$, 
leaving the structure of the equations unaltered. Choosing the $\a$ 
states such that
\begin{equation}
(H_{0}+W)_{\a\a'}=\d_{\a\a'}E'_{\a}
\end{equation}
the algebric system reduces to
\begin{equation}
\left( E_{m}-E\right) a_{m}+{\sum_{m^{\prime }}}a_{m^{\prime
}}W_{m,m^{\prime }}+{\sum_{\alpha }}b_{\alpha }W_{m,\alpha }=0 
\label{reson1}
\end{equation}
\begin{equation}
\left( E'_{\alpha }-E\right) b_{\alpha }+{\sum_{m^{\prime }}}a_{m^{\prime
}}W_{\alpha ,m^{\prime }}=0.  
\label{reson2}
\end{equation}
Solving for $b_{\alpha }$ and substituting in the first equation we exactly
decouple the 4-body states as well, ending up with an 
equation for the dressed pair $|a\rangle=\sum_{m}a_{m}|m\rangle$. 
The effective Schr\"{o}dinger equation
for the pair reads
\begin{equation}
\left( H_{0}+W+W_{eff}[E]\right) |a\ket =E|a\ket  
\label{can1}
\end{equation}
where
\begin{equation}
   (W_{eff})_{mm'}=-\sum_{\a}\frac{W_{m\a}W_{\a m'}}{E'_{\a}-E}.
\end{equation}
Hence Eq.(\ref{can1}) is a self-consistent equation and 
we have calculated the effective interaction $W_{eff}$ between 
$W=0$ pairs analytically\cite{cbs2},\cite{cbs3}; it can be attractive 
depending on $n$. Also we have  found that the results 
compare well with exact diagonalization results, when available. 
Basically the same mechanism works for small clusters with open 
boundary conditions\cite{cibal}. 

We want to stress that the truncated expansion of $|\Q_{0}(n+2)\ket$ in Eq.(\ref{lungo}) 
doesn't give a good approximation of the interacting ground state 
wave function but only of its weak 
coupling $a_{m}$ amplitudes.
Similarly in the BCS model, 
from the Cooper equation (obtained by truncation of $|\Q_{0}(n+2)\ket$ to the $m$ 
states) we can estimate only the pair coefficients of the ground state 
and not its full structure. Nevertheless this is enough to study 
bound states formation; indeed the energy gap 
of the pair in the Cooper theory and in the 
many-body BCS theory are equals.

\subsection{Truncated Configuration Interaction: Description of the 
Pairing Mechanism in Small Clusters}
\label{smallcluster}

Although the above canonical transformation can be performed without any truncation 
of the configuration interaction expansion, the description 
simplifies if we take into account $\a$ states only. As observed above 
this is justified in the weak coupling regime and is enough to study 
the pairing problem. Let us consider 
for example the pairing mechanism in small clusters, like $CuO_{4}$ and 
$Cu_{5}O_{16}$. Such clusters with 4 holes 
have ground states of 
$^{1}B_{2}$ symmetry in $C_{4v}$ for $0< U\leq t$. The diagnosis that hole pairing 
between the holes of a $W=0$ pair 
occurs in these ground states results from the following steps.

$a$) The lowest eigenstate of the one-body energy spectrum has $A_{1}$ 
symmetry and the interacting ground state with two holes is a $C_{4v}$ 
totalsymmetric singlet. The state 
$|\F\ket=c^{\dag}_{A_{1}\ua}c^{\dag}_{A_{1}\da}|0\ket$ is the 
two-body $U=0$ non-degenerate ground state, where here and in the 
following steps $c^{\dag}_{I\s}$ is the creation operator for the hole 
eigenstate of $I$ symmetry and spin $\s$. $b$) The first excited level 
of the one-body energy spectrum is degenerate and the corresponding 
eigenstates transform as the vector components $(x,y)$. $c$) To have 
a ground state with four holes of $^{1}B_{2}$ symmetry in the limit 
of vanishing interaction we must create two holes over $|\F\ket$ in 
the singlet state $|m_{S}\ket\equiv \frac{1}{\sqrt{2}}(c^{\dag}_{x\ua}c^{\dag}_{y\da}+
c^{\dag}_{y\ua}c^{\dag}_{x\da})
|\F\ket$. $d$) Taking $|m_{S}\ket$ as our unique $m$ state we apply 
the above canonical transformation and we find an effective attractive 
interaction summing over all the virtual 4-body (3 holes-1 electron) 
intermediate states. 

As another description to this weak-coupling 
case, we may say we are doing degenerate second-order perturbation 
theory; in this example, the zero-th approximation eigenstates are just 
two, $|m_{1}\ket\equiv c^{\dag}_{x\ua}c^{\dag}_{y\da}|\F\ket$ and 
$|m_{2}\ket\equiv c^{\dag}_{y\ua}c^{\dag}_{x\da}|\F\ket$, 
and the problem is reduced to a $2 \times 2$ matrix diagonalization. 
The corresponding second-order interaction is illustrated in Figure I: 
it  takes the state $|m_{1}\ket$ into $|m_{2}\ket$ and conversely. Hence  
it is actually a spin-flip interaction.
\begin{center}
 \begin{picture}(100,100)(0,0)
\SetOffset(-40,0)
\Text(20,20)[r]{$x\ua$}\Text(20,70)[r]{$y\da$}
\Text(170,45)[l]{$y\ua$}\Text(170,70)[l]{$x\da$}
\dl(70,45)(70,70){2}\ve(70,45){2.5}\ve(70,70){2.5}
\al(20,70)(70,70)\al(70,70)(120,70)\al(120,70)(170,70)
\al(70,45)(120,45)\al(120,45)(170,45)
\dl(120,20)(120,70){2}\ve(120,20){2.5}\ve(120,70){2.5}
\al(20,20)(120,20)\al(120,20)(70,45)
\end{picture}
\end{center}
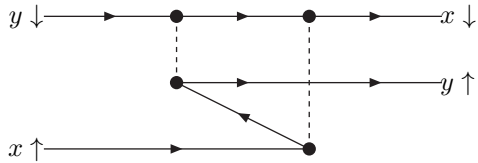
Figure I. {\small {\it The e-h exchange diagram for the Two-Hole Amplitude. 
For $W=0$ pairs, the direct interaction vanishes and this diagram 
produces an effective interaction, splitting singlet and triplet pairs.}}
\vspace*{0.5cm}

It turns out \cite{cibal} 
that the effective interaction is attractive in the singlet
$|\Psi_{^{1}B_{2}}\ket\equiv|m_{S}\ket=\frac{1}{\sqrt{2}}(|m_{1}\ket+|m_{2}\ket)$ 
and equal in size, but repulsive, in the triplet
$|\Psi_{^{3}B_{2}}\ket\equiv |m_{T}\ket=\frac{1}{\sqrt{2}}(|m_{1}\ket-|m_{2}\ket)$. 
That is definitely a pairing situation for the singlet. 
$e$) The two-body state $\frac{1}{\sqrt{2}}
(c^{\dag}_{x\ua}c^{\dag}_{y\da}+
c^{\dag}_{y\ua}c^{\dag}_{x\da})|0\ket$ is a $W=0$ pair (no direct  
interaction occurs between the holes); hence we can say that the $W=0$ 
pair, when dressed by virtual e-h excitations, gives up a bound pair.

These are  the principles  that here we wish to extend to the half 
filled $4\times 4$ Hubbard model of Eq.(\ref{hamil}).

\subsection{Plan of the paper}
\label{plane}

Below, we  consider the ground state of the $4 \times 4$ 
cluster with 14, 15 and 16 holes, and demonstrate that the 14 hole 
case must be interpreted in terms of pairing of two electrons added 
to the half-filled antiferromagnet. However the non-interacting ground 
state is degenerate and this means that we must extend the above formalism.

The physical interpretation emerges from the fact that we are able to 
solve the problem analytically at weak coupling. When the analytic 
results are compared to the numerical ones\cite{fop} we can conclude that 
$i$) we are able to predict from Group Theory the good quantum numbers and degeneracies of 
the ground states involved at various occupancies $ii$) we are able to show that 
the effective interaction between the extra electrons is attractive in 
the singlet (ground state) and repulsive for the triplet. 

These analytic results have been allowed by a deeper symmetry analysis 
of the $4 \times 4$ cluster than had been possible previously. The 
antiferromagnetic ground state has also been explored analytically by 
a new approach. In this way, the electron pair creation is 
also accoplished analytically, with full control of the symmetry.

Due to the special role played by $W=0$ pairs, in the next Section we state a 
general theorem to obtain all the possible $W=0$ pairs. The theorem 
requires the knowledge of a (previously unknown) symmetry Group big enough to explain the 
one-body  degeneracies (Optimal Group). As we shall see 
the Space Group ${\mathbf G}$ does not work and in Section 
\ref{optimal} we determine the Optimal Group for the $4\times 
4$ square lattice. In Section \ref{localform} we 
determine the exact weak coupling ground state wave function at half 
filling, that is 
unique as granted by the Lieb's theorem\cite{l}. This will be done using a 
new {\em local} formalism that enables us to write down all the $W=0$ 
states  of the partially filled shell. 
In Section \ref{12h}, with the help of the theorem 
of Section \ref{useful} and with the Optimal Group in hands, we single out all $W=0$ pairs 
formed from degenerate orbitals of the shell $\e(k)=0$. Hence in 
Section \ref{quartetti} we use the corresponding local basis to write 
down $W=0$ two and four-body states.  Then to study the 
pairing problem at half filling we extend the above canonical 
transformation and finally in Section \ref{mechanism} 
we shall reduce the pairing problem to $2 \times 2$ matrices 
and compare with the literature numerical data. Finally, we underline 
the implications  of the present results 
in Section \ref{conc}.

\section{W=0 pairs and symmetries of the doped ground states: a 
useful theorem}
\label{useful}

In general, one has a set ${\cal S}$
of degenerate eigenstates of $H_{0}$ which is partially filled in 
the $U=0$ limit. To first-order, we ignore the particles in the filled 
shells and  find the {\it exact} ground state(s) 
of $W$ in the truncated Hilbert space ${\cal H }$ describing those in 
the partially filled shell. One simple case occurs when only two 
holes are in the partially filled shell, because the ground states are $W=0$ 
pairs (filled shells are understood). We recall that they are two body singlets that are 
eigenstates of $H_{0}$ and belong to the kernel of $W$. There 
is no double occupation of any site in such a pair.
This means that the two 
particles of a $W=0$ pair do not interact directly, but only by means 
of virtual electron-hole excitations. 

As we shall see, another special case occurs when the system at half filling is 
doped by  removing two particles (they can only be removed as $W=0$ 
pairs).

In this Section we want to show how $W=0$ pairs arise, due to the 
symmetry of the system. Previously\cite{cbs1}\cite{cbs2} 
we have shown that $W=0$ pairs with zero 
total momentum are a consequence of the $C_{4v}$ point symmetry.
Projecting the determinantal state
\begin{equation}
|d(k)\ket=c^{\dag}_{k\ua}c^{\dag}_{-k\da}|0\ket
\end{equation}
on the irreducible representations  $A_{2}$, $B_{1}$ and $B_{2}$ 
of $C_{4v}$ one obtains $W=0$ pairs. 

Here we want to point out a  more powerful and elegant criterion to get 
\underline{all} the $W=0$ pairs, including those of nonvanishing  
total  momentum. We can do that in terms of the Optimal Group  ${\cal G}$ 
of the Hamiltonian, that we define as a symmetry Group which is 
big enough to justify the degeneracy of 
the single particle energy levels.  By definition, every one-body eigenstate of $H$ can be 
classified as belonging to one of the irreps of ${\cal G}$. We may 
say that an irrep $\eta$ is represented in the one-body spectrum of $H$ if 
at least one of the one-body levels belongs to $\eta$. Let
${\cal E}$ be the set of the irreps of ${\cal G}$ which are 
represented in the one-body spectrum of $H$. Let $|\q\rangle$ be a two-body eigenstate of 
the kinetic energy $ H_{0}$ with spin $ S_{z}=0$ and $P^{(\eta)}$ the 
projection operator on the irrep $\eta$. We wish to prove the 

{\it W=0 Theorem:
\begin{equation}
\eta \notin  {\cal E} \Leftrightarrow W P^{(\eta)}|\q\rangle=0 
\label{theo}
\end{equation}
In other terms, any nonvanishing projection of $ |\q\rangle$ on an irrep
\underline{not} contained in ${\cal 
E}$, is an eigenstate of $ H_{0}$ 
with no double occupancy. The singlet component of this state is a  
$W=0$ pair. Conversely, any pair belonging to an irrep represented in the 
spectrum must have positive $W$ expectation value}. 

In the case of the $4\times 4$ model (see Section \ref{12h}) the pairs 
belonging to irreps of the Optimal Group have well defined 
parities under particle exchange, and can be classified as singlet or 
triplet pairs.

{\it Proof}: Let us consider a  
two body state of opposite spins trasforming 
as the $i$-th component of the irrep $\eta$ of ${\cal G}$:
\begin{equation}
|\q_{i}^{(\eta)}\rangle=\sum_{r_{1}r_{2}}\q_{i}^{(\eta)}(r_{1},r_{2})
    c^{\dag}_{r_{1}\ua}c^{\dag}_{r_{2}\da}|0\rangle.
\end{equation}
Then we have
\begin{equation}
 \hat{n}_{r\ua}\hat{n}_{r\da}|\q_{i}^{(\eta)}\rangle=
 \q_{i}^{(\eta)}(r,r)c^{\dag}_{r\ua}c^{\dag}_{r\da}|0\rangle\equiv 
    \q_{i}^{(\eta)}(r,r)|r\ua,r\da\rangle.
\end{equation}
We define $P_{i}^{(\eta)}$ as the projection operator on the $i$-th 
component of the irrep $\eta$. Since
\begin{equation}
   P_{i}^{(\eta)}\sum_{r}\q_{i}^{(\eta)}(r,r)|r\ua,r\da\rangle=
   \sum_{r}\q_{i}^{(\eta)}(r,r)|r\ua,r\da\rangle
\end{equation}
if
\begin{equation}
 P_{i}^{(\eta)}|r\ua,r\da\rangle=0\;\;\;\;\;\forall r
\label{w=0cond}
\end{equation}
then $\q_{i}^{(\eta)}(r,r)=0\;\forall r$.
It is worth to note that Eq.(\ref{w=0cond}) is true if and only if
\begin{displaymath}
    P_{i}^{(\eta)}|r\s\rangle=0\;\;\;\;\;\forall r
\end{displaymath}
where $|r\s\rangle=c^{\dag}_{r\s}|0\ket$. 

It is always possible to 
write $ |r\s\rangle$ as 
\begin{equation}
   |r\s\rangle=\sum_{\eta\in {\cal E}}\sum_{i}c^{(\eta)}_{i}(r)
    |\vf^{(\eta)}_{i,\s}\rangle
    \label{rdecomp}
\end{equation}
where $|\vf^{(\eta)}_{i,\s}\rangle$ is the one-body eigenstate of $H_{0}$ 
with spin $\s$ transforming as the $i$-th component of the irrep 
$\eta$.  From (\ref{rdecomp}) it follows directly that if 
$ \eta'$ does not belong to ${\cal E}$
\begin{displaymath}
 P^{(\eta')}|r\s\rangle=0 
\end{displaymath}
and so $P^{(\eta')}|r\ua,r\da\rangle=0$. $\Box$ 

This theorem restricts the possible ground state symmetries in some 
special case. Let $U_{c}(n)$ be the minimum crossover value of $U$, 
that is the ground state with $n$ holes $|\Q_{0}(n)\ket$ has well defined symmetry 
$\eta_{0}$ for $0<U<U_{c}(n)$. Let $|\F(n)\ket$ be non-degenerate 
(closed shells case). Then if we add two extra 
holes, the new ground state for $U=0$ is a $W=0$ pair over 
$|\F(n)\ket$, and its symmetry is $\eta_{0}\cdot\eta_{W=0}$, where 
$\eta_{W=0}$ is the symmetry of the added pair. Turning on the 
interaction the symmetry cannot change if $U<U_{c}(n+2)$! This 
restriction is posed by Group theory alone: which of the symmetries 
that remain allowed is actually realised in the ground state depends 
on the dynamics. 

The complete characterization of the symmetry of $W=0$ pairs requires 
the knowledge of the Optimal Group ${\cal G}$. A partial use of the 
theorem is possible if one  does not know 
${\cal G}$ but knows a subgroup, like the Space Group  $ \mathbf{G}$.
It is then still granted that any pair belonging to an irrep of $ \mathbf{G}$ not
represented in the spectrum has the $W=0$ property.  On the other 
hand, accidental degeneracies occur with a subgroup of the Optimal 
Group, and by mixing degenerate pairs belonging to irreps represented 
in the spectrum one can find $W=0$ pairs also there. This is clearly 
illustrated by the example repported in the next Section.

\section {The Optimal Group for the $4 \times 4$ Hubbard Model}
\label{optimal}

The half filled shell of the $4 \times 4$ Hubbard model 
has degeneracy 6, and the Space Group 
$\mathbf{G}$ does not have irreps with dimensions bigger than 
4\cite{hamer}. An additional symmetry is needed to justify the sixfold 
degenerate  eigenvalue $\e(k)=0$, and we have found it\cite{ijmp}.
Let us represent the $4\times 4$ lattice as 
\begin{center}
    \vspace*{0.5cm}
     \begin{tabular}{|c|c|c|c|}
\hline
   1 & 2  & 3 & 4 \\
   \hline
   5 & 6 & 7 & 8 \\
   \hline
   9 & 10 & 11 & 12 \\
   \hline
   13 & 14 & 15 & 16\\
   \hline
   \end{tabular}
   \vspace*{0.5cm}
\end{center}
Periodic boundary conditions are assumed and  for 
example, the nearest neighbours of 1 are 2, 5, 4 and 13. Rotating 
 the plaquettes 1,2,5,6 and 11,12,15,16 clockwise and the other two 
counterclockwise  by 90 degrees we obtain the effect of the 
``dynamical'' symmetry, that we call $d$:
\begin{center}
    \vspace*{0.5cm}
     \begin{tabular}{|c|c|c|c|}
\hline
   5 & 1  & 4 & 8 \\
   \hline
   6 & 2 & 3 & 7 \\
   \hline
   10 & 14 & 15 & 11 \\
   \hline
   9 & 13 & 16 & 12\\
   \hline
   \end{tabular}
   \vspace*{0.5cm}
\end{center}
This transformation preserves nearest neighbours (and so, each order
of neighbours) but  is not an isometry, and for example the distance 
between 1 and 3 changes.  Thus, this symmetry operation $d$ is a new,
dynamical symmetry. Including  $d$ and closing the multiplication table
we obtain the Optimal Group ${\cal G}$  with 384 elements in 
20 classes (like $\mathbf{G}$) as shown in Table I.
\begin{center}
    \vspace*{0.5cm}
     \begin{tabular}{|c|c|c|c|c|c|c|c|c|c|}
	\hline
	${\cal C}_{1}$&${\cal C}_{2}$&${\cal C}_{3}$&${\cal C}_{4}$&${\cal C}_{5}$&
	${\cal C}_{6}$&${\cal C}_{7}$&${\cal C}_{8}$&${\cal C}_{9}$&${\cal C}_{10}$\\
	\hline
	$I$&$t_{22}$&$C_{4}\sigma'[2]$&$\sigma_{x}$&$C_{2}$&$\sigma'$&$C_{2}d$
	&$C_{2}t_{22}d$&$C_{2}\sigma'[1]$&$C_{4}^{3}t_{02}d$\\
	\hline
	${\cal C}_{11}$&${\cal C}_{12}$&${\cal C}_{13}$&${\cal C}_{14}$&${\cal C}_{15}$
	&${\cal C}_{16}$&${\cal C}_{17}$&${\cal C}_{18}$&${\cal C}_{19}$&${\cal C}_{20}$\\
	\hline
	$C_{4}^{3}t_{20}d$&$C_{2}t_{01}$&$C_{2}\sigma_{x}[1]$&$C_{4}$&$C_{2}t_{01}d$&$C_{2}t_{12}d$
	&$C_{2}\sigma_{x}[1]d$&$C_{4}[1]d$&$C_{2}[1]d$&$C_{4}[1]$\\
	\hline
\end{tabular}
   \vspace*{0.5cm}
\end{center}
Table I. {\small {\it Here, we report one operation for each of the 20 classes ${\cal C}_{i}$;
the others can be obtained by conjugation. The operations 
are: the identity $I$, the translation $t_{mn}$ of $m$ steps along $x$ 
and $n$ along $y$ axis; the 
other operations $C_{2},C_{4},\sigma,\sigma'$ are those of the Group of 
the square and are referenced to the centre;
however,$C_{2}[i],C_{4}[i],\sigma[i],$ and $\sigma'[i]$ are 
centered on site $i$.}}
\vspace*{0.5cm}

The complete  Character  Table of  ${\cal G}$  is shown as Table II.
\begin{center}
    \vspace*{0.5cm}
     \begin{tabular}{|c|c|c|c|c|c|c|c|c|c|c|c|c|c|c|c|c|c|c|c|c|}
	\hline
	${\cal G}$&${\cal C}_{1}$&${\cal C}_{2}$&${\cal C}_{3}$&${\cal C}_{4}$&
	${\cal C}_{5}$&${\cal C}_{6}$&${\cal C}_{7}$
	&${\cal C}_{8}$&${\cal C}_{9}$&${\cal C}_{10}$&${\cal C}_{11}$&${\cal C}_{12}$
	&${\cal C}_{13}$&${\cal C}_{14}$&${\cal C}_{15}$
	&${\cal C}_{16}$&${\cal C}_{17}$&${\cal C}_{18}$&${\cal C}_{19}$&${\cal C}_{20}$\\
	\hline
        $A_{1}$&1&1&1&1&1&1&1&1&1&1&1&1&1&1&1&1&1&1&1&1\\
	\hline
        $\Tilde A_{1}$&1&1&1&1&1&1&-1&-1&1&-1&-1&-1&-1&1&1&1&1&1&-1&-1\\
	\hline
	$B_{2}$&1&1&-1&-1&1&1&1&1&1&-1&-1&1&-1&-1&1&1&-1&-1&1&-1\\
	\hline
	$\Tilde B_{2}$&1&1&-1&-1&1&1&-1&-1&1&1&1&-1&1&-1&1&1&-1&-1&-1&1\\
	\hline
	$\G_{1}$&2&2&-2&-2&2&2&0&0&2&0&0&0&0&-2&-1&-1&1&1&0&0\\
	\hline
	$\G_{2}$&2&2&2&2&2&2&0&0&2&0&0&0&0&2&-1&-1&-1&-1&0&0\\
	\hline
	$\S_{1}$&3&3&3&3&3&-1&-1&-1&-1&-1&-1&-1&-1&-1&0&0&0&0&1&1\\
	\hline
	$\S_{2}$&3&3&3&3&3&-1&1&1&-1&1&1&1&1&-1&0&0&0&0&-1&-1\\
	\hline 
	$\S_{3}$&3&3&-3&-3&3&-1&-1&-1&-1&1&1&-1&1&1&0&0&0&0&1&-1\\
	\hline
	$\S_{4}$&3&3&-3&-3&3&-1&1&1&-1&-1&-1&1&-1&1&0&0&0&0&-1&1\\
	\hline
	$\L_{1}$&4&-4&-2&2&0&0&-2&2&0&-2&2&0&0&0&-1&1&-1&1&0&0\\
	\hline 
	$\L_{2}$&4&-4&-2&2&0&0&2&-2&0&2&-2&0&0&0&-1&1&-1&1&0&0\\
	\hline 
	$\L_{3}$&4&-4&2&-2&0&0&-2&2&0&2&-2&0&0&0&-1&1&1&-1&0&0\\
	\hline
	$\L_{4}$&4&-4&2&-2&0&0&2&-2&0&-2&2&0&0&0&-1&1&1&-1&0&0\\
	\hline
	$\W_{1}$&6&6&0&0&-2&-2&-2&-2&2&0&0&2&0&0&0&0&0&0&0&0\\
	\hline
	$\W_{2}$&6&6&0&0&-2&-2&2&2&2&0&0&-2&0&0&0&0&0&0&0&0\\
	\hline
	$\W_{3}$&6&6&0&0&-2&2&0&0&-2&-2&-2&0&2&0&0&0&0&0&0&0\\
	\hline
	$\W_{4}$&6&6&0&0&-2&2&0&0&-2&2&2&0&-2&0&0&0&0&0&0&0\\
	\hline
	$\P_{1}$&8&-8&-4&4&0&0&0&0&0&0&0&0&0&0&1&-1&1&-1&0&0\\
	\hline
	$\P_{2}$&8&-8&4&-4&0&0&0&0&0&0&0&0&0&0&1&-1&-1&1&0&0\\
	\hline
\end{tabular}
\vspace*{0.5cm}

Table II. {\small {\it Character Table of the {\em Optimal Group} ${\cal G}$ of the 
$4 \times 4$ model.}}
\end{center}
\vspace*{0.5cm}

As the notation suggests, the irreps $A_{1}$ and $\tilde A_{1}$ both 
reduce to  $A_{1}$, in $C_{4v}$, while  $B_{2}$ and $\tilde B_{2}$ both 
reduce to  $B_{2}$. Table III shows how the irreps of ${\cal G}$ 
split in $C_{4v}$.
\begin{center}
    \vspace*{0.5cm}
     \begin{tabular}{|c|c|}
	\hline
	${\cal G}$&$C_{4v}$\\
	\hline
        $A_{1}$&$A_{1}$\\
	\hline
        $\Tilde A_{1}$&$A_{1}$\\
	\hline
	$B_{2}$&$B_{2}$\\
	\hline
	$\Tilde B_{2}$&$B_{2}$\\
	\hline
	$\G_{1}$&$2B_{2}$\\
	\hline
	$\G_{2}$&$2A_{1}$\\
	\hline
	$\S_{1}$&$A_{1}+2B_{1}$\\
	\hline
	$\S_{2}$&$A_{1}+2B_{1}$\\
	\hline 
	$\S_{3}$&$2A_{2}+B_{2}$\\
	\hline
	$\S_{4}$&$2A_{2}+B_{2}$\\
	\hline
	$\L_{1}$&$A_{1}+B_{1}+E$\\
	\hline 
	$\L_{2}$&$A_{1}+B_{1}+E$\\
	\hline 
	$\L_{3}$&$A_{2}+B_{2}+E$\\
	\hline
	$\L_{4}$&$A_{2}+B_{2}+E$\\
	\hline
	$\W_{1}$&$A_{2}+B_{1}+2E$\\
	\hline
	$\W_{2}$&$A_{2}+B_{1}+2E$\\
	\hline
	$\W_{3}$&$A_{1}+B_{2}+2E$\\
	\hline
	$\W_{4}$&$A_{1}+B_{2}+2E$\\
	\hline
	$\P_{1}$&$2A_{1}+2B_{1}+2E$\\
	\hline
	$\P_{2}$&$2A_{2}+2B_{2}+2E$\\
	\hline
\end{tabular}
   \vspace*{0.5cm}
   
Table III. {\small {\it Reduction of the irreps of  Optimal Group ${\cal G}$ of the 
   $4\times 4$ model in the point Group.}}
\end{center}
We call ${\cal G}$ the Optimal Group because it enables us to explain
the degeneracy of the one-particle energy spectrum; in other terms, no 
accidental degeneracy of orbitals occours using ${\cal G}$. 
In the Table IV below we report the one-body eigenvalues for $t=-1$,
the degeneracy and the symmetry of each eigenvector. Below, we shall 
find that ${\cal G}$ is also adequate to classify the many-body ground 
states. For pairs, the {\it $W=0$ Theorem} ensures that no double 
occupancy is possible in the irreps $\Tilde 
A_{1},B_{2},\G_{1},\G_{2},\S_{1},\S_{2},\S_{3},\S_{4},\L_{2},\L_{3},\W_{1},
\W_{2},\W_{3},\P_{1}$ and $\P_{2}$.
\begin{center}
    \vspace*{0.5cm}
     \begin{tabular}{|c|c|c|}
\hline
   Energy&Irrep of ${\cal G}$ &Degeneracy \\
   \hline
   4 & $\Tilde B_{2}$ & 1 \\
   \hline
   2 & $\L_{4}$ & 4  \\
   \hline
   0 & $\W_{4}$ & 6 \\
   \hline
   -2 & $\L_{1}$ & 4 \\
   \hline
   -4 & $A_{1}$ & 1 \\
   \hline
   \end{tabular}
   \vspace*{0.5cm}
   
Table IV. {\small {\it One-body spectrum for $t=-1$.}}
\end{center}

\section{{\em Local} Formalism at Half Filling}
\label{localform}

As observed in the previous Sections, $W=0$ pairs predict the possible 
ground state symmetries of systems 
which differ from closed shells by a pair. Let us now consider how the 
above analysis extends to the doped half filled system. 
Let ${\cal S}_{hf}$ denote the set (or shell) of the $k$ wave vectors 
such that $\e(k)=0$. At half filling ($N^{2}$ holes) for $U=0$ the 
ground state has the ${\cal S}_{hf}$  shell 
half occupied, while all $|k\ket$ orbitals such that 
$\e(k)<0$ are filled. The $k$ vectors of ${\cal S}_{hf}$ lie on the square having 
vertices $(\pm\pi,0)$ and $(0,\pm\pi)$;
one  readily realizes that the dimension of the 
set ${\cal S}_{hf}$, is $|{\cal S}_{hf}|=2N-2$.

For $N=4$, the  6 wave vectors are $k_{1}=(\p,0),\;k_{2}=(0,\p),\;
k_{3}=(\p/2,\p/2),\;k_{4}=(\p/2,-\p/2),\;k_{5}=(-\p/2,-\p/2)$ and 
$k_{6}=(-\p/2,\p/2)$.
\begin{center}
 \begin{picture}(200,200)(0,0)
\SetOffset(-0,0)
\Line(20,20)(20,180)\Line(20,180)(180,180)\Line(180,180)(180,20)
\Line(180,20)(20,20)
\Text(160,104)[r]{$k_{1}$}\Text(100,160)[r]{$k_{2}$}\Text(135,120)[r]{$k_{3}$}
\Text(135,80)[r]{$k_{4}$}\Text(75,120)[r]{$k_{6}$}\Text(75,80)[r]{$k_{5}$}
\Text(205,100)[r]{($\p$,0)}\Text(100,201)[t]{(0,$\p$)}
\DashLine(100,180)(180,100){5}
\DashLine(100,20)(180,100){5}\DashLine(20,100)(100,20){5}\DashLine
(20,100)(100,180){5}
\al(100,100)(180,100)\al(100,100)(100,180)\al(100,100)(140,140)
\al(100,100)(60,60)\al(100,100)(60,140)\al(100,100)(140,60)
\label{brillouin}
\end{picture}
\end{center}
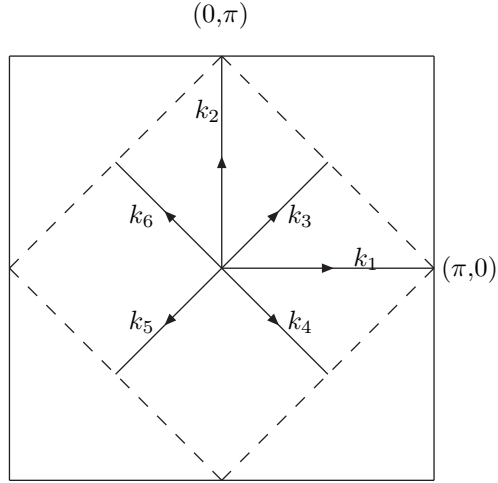
Figure I. {\small {\it The Brillouin Zone; the dashed 
square marks the condition of vanishing kinetic energy. 
${\cal S}_{hf}$ contains 6 states (arrows)  belonging to  
one-dimensional irreps of $\mathbf{T}$. Moreover, the arrows are mixed 
by operations of $C_{4v}$, and $(\pi,0)$ and $(0,\pi)$ 
 are the basis of a two-dimensional irrep of $\mathbf{G}$, while
the  wavevectors $k=(\pm\pi/2,\pm\pi/2)$  mix among themselves and yield
a four-dimensional irrep. The degeneracy of these two irreps is 
accidental for  $\mathbf{G}$, but is explained by the Optimal Group 
${\cal G}$. }}
\vspace{1cm}

Since $H$ commutes with the total spin operators,
\begin{equation}
\hat{S}_{z}=\frac{1}{2}\sum_{r}(\hat{n}_{r\ua}-\hat{n}_{r\da}),\;\;\;\;
\hat{S}^{+}=\sum_{r}c^{\dag}_{r\ua}c_{r\da},\;\;\;\;
\hat{S}^{-}=(\hat{S}^{+})^{\dag},
\label{su2gen}
\end{equation} 
at half filling every ground state of $H_{0}$ is represented in 
the $S_{z}=0$ subspace. Thus, $H_{0}$ has  
$\left(\begin{array}{c} 6 \\ 3 
\end{array}\right)^{2}$   degenerate unperturbed ground state 
configurations with $S_{z}=0$. We will show how this degeneracy is removed by the 
Coulomb interaction $W$ already in first-order perturbation theory. 
Actually most of the degeneracy is removed in first-order, and  with the help of 
Lieb's theorem\cite{l} we shall be  able to single out the true, unique first 
order ground state of $H$. In Appendix \ref{equivalence} we show that 
the structure of the first-order wave functions 
is gained by  diagonalizing $W$ in the {\bf  truncated  Hilbert space ${\cal H}$} 
spanned by the {\bf states of 3 holes of each spin in ${\cal S}_{hf}$}.
In other terms, one solves a 6-particle problem in the truncated  
Hilbert space ${\cal H}$ and then, understanding the particles in the 
filled shells, obtains the first-order ground state eigenfunction of $H$ in the 
full 16-particle problem. We underline that the matrix of $H_{0}$ 
in ${\cal H}$ is null,  since by construction ${\cal H}$ is contained in the 
kernel of $H_{0}$.

The operator $\sum_{r}\hat{n}_{r\ua}\hat{n}_{r\da}$ has 
eigenvalues $0,1,2, \ldots$ and so the lowest eigenvalue  of $W$ is zero 
(in other terms, $W$ is positive semi-definite). The  unique
ground state of the Hubbard Hamiltonian for $U=0^{+}$ at half
filling will turn out to be a $W=0$ singlet state of 6 holes in ${\cal S}_{hf}$ (filled 
shells being understood).  We shall obtain the  $W=0$ states $\in{\cal H}$. 
It is clear that, although the $U=0$ case is trivial, at  $U=0^{+}$ we are still 
facing a {\em bona fide}  many-body problem, that we are solving 
exactly\cite{he}.
In the present Section we define a basis of {\em local}  orbitals 
for  the $4 \times 4$ Hubbard model with periodic boundary conditions;
this basis is crucial for making the problem tractable,
both at half filling and for the doped 
case.  Using the local basis, 
the many body wave function of the antiferromagnetic ground state can 
be projected out as the singlet component of a {\bf single} determinant, 
which is amazingly simple for an interacting system; the effective 
interaction between the doped holes also emerges analytically.
The treatment for the half-filled case has already been generalized\cite{antif} 
to the $N \times N$ Hubbard model; below, we present 
the much simpler solution of the  $4 \times 4$ case, which is 
sufficient for our present purposes.

Since $W$ depends on the occupation number operators $\hat{n}_{r}$, 
it is intuitive that  its 
properties in ${\cal H}$ are best discussed by a suitable one-body 
basis of ${\cal S}_{hf}$ such that at 
least one of these operators is diagonal. In addition, a convenient basis 
should exploit the large ${\mathbf G}$ symmetry of the system. If  ${\cal S}_{hf}$ were a 
complete set ($N^{2}=16$ states), one would trivially go from plane waves 
to atomic orbitals by a Fourier transformation; instead,
we must  define the local counterparts of plane-wave states
using only the $2N-2=6$ states that belong to ${\cal S}_{hf}$. 

For each site $r$  we  diagonalize the number operator $\hat{n}_{r}$ 
(for the moment we omit the spin index);
it is a trivial matter to verify that  $(n_{r})_{ij}=\bra k_{i}|\hat{n}_{r}|
k_{j}\ket=\frac{1}{16}e^{i(k_{i}-k_{j})r}$ has eigenvalues $3/8$ and 
five times $0$.  This degeneracy suggests that we should diagonalize 
other operators in order to label the $\hat{n}_{r}$ eigenvactors, and indeed,
 since $\hat{n}_{r}$ is compatible with the operations of 
the point symmetry group $C_{4v}$ we also diagonalize the Dirac 
characters of this Group. The set of Dirac characters defines
the irreducible representation ({\it irrep}); thus we write the 
one-body basis states $\{|\varphi_{\a}^{(r)}\ket\}$ where  $\a$ comprises the $\hat 
{n}_{r}$ eigenvalue and an $C_{4v}$ irrep label. It is easy to verify that for $r=0$ the 
eigenvector with nonzero eigenvalue is just the totally symmetric 
superposition of all the  $\{|k_{i}\ket\} \in {\cal S}_{hf}$. 
Translating by $r$, plane wave states pick up a phase factor: 
$|k\ket\ra e^{ikr}|k\ket$. Hence the $\hat{n}_{r}$ eigenvector of occupation 
$3/8$ is
\begin{equation}
   |\phi_{1}^{(r)}\rangle\equiv |\vf^{(r)}_{A_{1}}\ket=
   \frac{1}{\sqrt{6}}\sum_{j=1}^{6}e^{ik_{j}r}|k_{j}\ket
    \label{occup}
\end{equation}
and we set up our local basis at $r$ by
\begin{equation}
   |\phi_{i}^{(r)}\rangle=\sum_{j=1}^{6}O_{ij}e^{ik_{j}r}|k_{j}\ket
    \label{orbi}
\end{equation}
\begin{equation}
    |k_{j}\ket=e^{-ik_{j}r}\sum_{n=1}^{6}O_{nj}|\phi_{n}^{(r)}\rangle
\label{inverso}
\end{equation}
where we introduce the orthogonal matrix
\begin{equation}
O=\frac{1}{\sqrt{6}}
\left[\begin{array}{rrrrrrrr}
                              1 & 1 & 1 & 1 & 1 & 1 \\
                              \sqrt{2} & \sqrt{2} & \frac{-1}{\sqrt{2}} &\frac{-1}{\sqrt{2}} & \frac{-1}{\sqrt{2}} & \frac{-1}{\sqrt{2}} \\
			      0 & 0 & \sqrt{\frac{3}{2}} & -\sqrt{\frac{3}{2}} & \sqrt{\frac{3}{2}} & -\sqrt{\frac{3}{2}} \\
			       \sqrt{3} & -\sqrt{3} & 0 & 0 & 0 & 0 \\
			       0 & 0 & \sqrt{3} & 0 & -\sqrt{3}& 0 \\
			       0 & 0 & 0 & \sqrt{3} & 0 & -\sqrt{3} 
\end{array}\right]
\label{ort6}
\end{equation}
It is clear from Eqs.(\ref{orbi})(\ref{ort6}) that
$|\phi_{i}^{(r)}\rangle$ has well defined occupation $n$ and symmetry
for point group 
operations centered at site $r$;  namely, it has $n=3/8$ for $i=1$ 
and $n=0$ otherwise;
it belongs to $A_{1}$ for 
$i=1,2$, to $B_{2}(xy)$ for $i=3$, to $B_{1}(x^{2}-y^{2})$ for $i=4$ 
and to $E$ for $i=5,6$. The local bases of different sites $r$ and 
$r\prime$ are 
connected by the unitary  transformation
\begin{equation}
|\phi_{i}^{(r)}\rangle=\sum_{j=1}^{6}|\phi_{j}^{(r^{\prime})}\rangle 
T^{(r^{\prime},r)}_{j,i}  
\label{ovvia}
\end{equation}
and using the orthonormality of the $|k\ket$ states we obtain the 
elements of the symmetric {\em translation} matrix
\begin{equation}
    T^{(r^{\prime},r)}_{j,i}=\langle 
    \phi_{j}^{(r^{\prime})}|\phi_{i}^{(r)}\rangle
    =\sum_{m=1}^{6}O_{j,m}O_{i,m}e^{ik_{m}.(r-r^{\prime})}   
\label{ritranslation}
\end{equation}
The translation  matrix {\em knows} all the ${\mathbf G}$  symmetry of the 
 system, and must be very special. Using such a basis set for the half 
filled shell the antiferromagnetic order of the ground state comes 
out in a clear and transparent manner. It is  clear that 
$[T^{(r^{\prime},r)}]^{4}=1$. The translation by one step towards the right 
is accomplished by
\begin{equation}
T^{(right)}=
\left[\begin{array}{rrrrrrrr}
                0 & 0 & 0 & \frac{-1}{\sqrt{3}} & \frac{i}{\sqrt{3}} & \frac{i}{\sqrt{3}} \\
                0 & 0 & 0 &-\sqrt{\frac{2}{3}} & \frac{-i}{\sqrt{6}} & \frac{-i}{\sqrt{6}} \\
	        0 & 0 & 0 & 0 & \frac{i}{\sqrt{2}} & \frac{-i}{\sqrt{2}} \\
               \frac{-1}{\sqrt{3}} & -\sqrt{\frac{2}{3}} & 0 & 0 & 0 & 0 \\
			       \frac{i}{\sqrt{3}} & \frac{-i}{\sqrt{6}} & \frac{i}{\sqrt{2}} & 
			       0 & 0& 0 \\
			       \frac{i}{\sqrt{3}} & \frac{-i}{\sqrt{6}} & \frac{-i}{\sqrt{2}} & 
			       0 & 0 & 0
\end{array}\right]
\label{translationx}
\end{equation}
The matrix that makes one step upwards is 
\begin{equation}
T^{(up)}=
\left[\begin{array}{rrrrrrrr}
                    0 & 0 & 0 & \frac{1}{\sqrt{3}} & \frac{i}{\sqrt{3}} & \frac{-i}{\sqrt{3}} \\
                    0 & 0 & 0 &\sqrt{\frac{2}{3}} & \frac{-i}{\sqrt{6}} & \frac{i}{\sqrt{6}} \\
	            0 & 0 & 0 & 0 & \frac{i}{\sqrt{2}} & \frac{i}{\sqrt{2}} \\
	            \frac{1}{\sqrt{3}} & \sqrt{\frac{2}{3}} & 0 & 0 & 0 & 0 \\
	            \frac{i}{\sqrt{3}} & \frac{-i}{\sqrt{6}} & \frac{i}{\sqrt{2}} & 0 & 0& 0 \\
	            \frac{-i}{\sqrt{3}} & \frac{i}{\sqrt{6}} & \frac{i}{\sqrt{2}} & 0 & 0 & 0
\end{array}\right]
\label{translationy}
\end{equation}

The reason why this choice of the basis set is clever is now 
apparent. The local basis at any site $r$ splits into the subsets
${\cal S}_{a}=\{|\phi_{1}^{(r)}\rangle,|\phi_{2}^{(r)}\rangle,
|\phi_{3}^{(r)}\rangle\}$, and
${\cal S}_{b}=\{|\phi_{4}^{(r)}\rangle,|\phi_{5}^{(r)}\rangle,
|\phi_{6}^{(r)}\rangle\}$;
a shift by a lattice step sends members of ${\cal S}_{a}$ into linear combinations 
of the menbers of  ${\cal S}_{b}$, and conversely.

Consider the 6-body determinantal eigenstate of $H_{0}$
\begin{equation}
  |d^{(r)}[1,2,3]\ket_{\s} =
  |\phi^{(r)}_{1,\s}\phi^{(r)}_{2,\s}\phi^{(r)}_{3,\s}
  \phi^{(r)}_{4,-\s}\phi^{(r)}_{5,-\s}\phi^{(r)}_{6,-\s}\ket ;
  \label{antidiag}
\end{equation}
the notation implies that $|d^{(r)}[i,j,k]\ket_{\s}$ denotes a 6-body 
determinant with one body per local state and $i,j,k$ with spin 
$\s$, the complement with spin $-\s$; local states are ordered in the 
natural way $1,\dots 6$. In this state there is partial occupation of 
site $r$ with spin $\s$, but no double occupation. Introducing the 
primitive translation of the lattice $\hat{e}_{x}=(1,0)$ and 
$\hat{e}_{y}=(0,1)$, it turns out that a shift by a lattice 
step $r\ra r'=r\pm\hat{e}_{l}$ with $l=x,y$, produces the transformation
\begin{equation}
  |d^{(r)}[1,2,3]\ket_{\s}\longleftrightarrow|d^{(r')}[1,2,3]\ket_{\s}
  =-|d^{(r)}[4,5,6]\ket_{\s},
\label{giochino2}
\end{equation}
that is, a lattice step is equivalent to a spin flip ({\em 
antiferromagnetic property}). Since the spin-flipped state is 
also free of double occupation, $ |d^{(r)}[1,2,3]\ket_{\s}$ is a $W=0$  
6-body eigenstate of $H$. A ground state 
which is a single determinant is a quite remarkable property of an 
interacting model like this, and this property holds at half 
filling, not in general. To be sure, $|d^{(r)}[1,2,3]\ket_{\s}$ is 
a mixture of  pure spin components $|\F_{AF}^{S}\ket$ with $S=0,1,2,3$. 
However, $W$ is positive semi-definite and this implies that 
all the pure spin components must possess the $W=0$ property as well. In 
particular, the singlet $|\F_{AF}^{S=0}\ket$ is a $W=0$ eigenstate and is the true ground state of 
the Hubbard model at half filling which is predicted by Lieb's 
theorem (filled shell are understood). 
Explicitly, the antiferromagnetic ground state wave function reads
\begin{equation}
|\F_{AF}^{S=0}\ket=\Hat{A} (3,6)\Hat{A} (2,5)\Hat{A} 
(1,4)|d^{(r)}[1,2,3]\ket_{\s},
\label{af}
\end{equation}
where $\Hat{A}$ is the antisimmetrizer, such that for example 
$\Hat{A}(1,4)|d^{(r)}[1,2,3]\ket_{\s}=|d^{(r)}[1,2,3]\ket_{\s}-
|d^{(r)}[4,2,3]\ket_{\s}$. One can easly verify that $|\F_{AF}^{S=0}\ket$ is 
independent by the $r$ and $\s$ label of $|d^{(r)}[1,2,3]\ket_{\s}$, 
modulo phase factors. 
From the general analysis of Ref.\cite{antif} we obtain that this 
singlet has $A_{1}$ symmetry with respect to the center of an 
arbitrary plaquette of the square lattice and vanishing total momentum 
as in the strong coupling limit\cite{md}.
It is worth noticing that the open-shell part of the antiferromagnetic ground state
(not considering the occupied inner shells) is a 
6-body $W=0$ singlet state. Correlation effects enable no fewer than 
6 particles to completely avoid double occupation in such a small 
system. This is also a consequence of Lieb's theorem. If all the 6 
body are taken with parallel spin, double occupation is trivially 
avoided; however, Lieb's theorem enforces a singlet ground state, so 
a singlet $W=0$ state must exist.

The $4 \times 4$ case at hand can be thorougly explored 
on the computer, since the size of  ${\cal H}$ at half filling is 400. 
We have used Mathematica to diagonalize $H+\xi S^{2}$, where a small 
$\xi$ is a numerical device to keep the different spin components of 
the ground state separated. In this way, we {\em observed} the fourfold 
degenerate, $W=0$ ground state which $\xi$ separates into its singlet, 
triplet, quinted and septet components, as expected. At $\x=0$ the 
separation grows like $U^{2}$. The 
antiferromagnetic property of the wave functions was also easily and 
nicely borne out by the numerical results.  

\section {$W=0$ Pairs and Quartets in the plane-wave representation.}
\label{12h}

In this Section we use the antiferromagnetic ground 
state $|\F_{AF}^{S=0}\ket$ to predict the possible symmetries 
of the doped half filled system.  
With 12 holes, in the $U \rightarrow 0$ limit, there are two in
${\cal S}_{hf}$; the first-order ground states correspond to $W=0$
pairs.  The symmetry of these $W=0$ states can be determined {\em a 
priori} from the ${\cal G}$  irreps of Table IV. Apart from the filled 
shells, two holes go to the $\Omega_{4}$ level (Table IV). 
From the character Table II one can derive that 
\begin{equation}
\Omega_{4}^{2}=A_{1}+\tilde{B}_{2}+\Omega_{4}+\Gamma_{1}+
\Gamma_{2}+\Sigma_{2}+\Sigma_{3}+\Omega_{1}+\Omega_{2}+\Omega_{3};
\label{quadrato}
\end{equation}
since the first 3 entries are present in Table IV, the {\it $W=0$  
Theorem} ensures that $\Gamma_{1},
\Gamma_{2},\Sigma_{2},\Sigma_{3},\Omega_{1},\Omega_{2}$ and
$\Omega_{3}$ pairs have no double occupation. It turns out that the 
spin and orbital symmetries are entangled, i.e. some of these pairs 
are triplet and the rest singlet.  We can see that by 
projecting the determinantal state 
$c^{\dag}_{k\ua}c^{\dag}_{p\da}|0\ket$ with $k,p\in 
{\cal S}_{hf}$ on the irreps not contained 
in the spectrum. One obtaines singlet $W=0$ pairs for 
$\G_{1},\;\G_{2},\;\S_{2},\;\W_{1}$; they read
\begin{equation}
    \begin{array}{l}
    |\q^{^{1}\G_{1}}_{1}\ket=\{\frac{1}{\sqrt{3}}
    (c^{\dag}_{k_{1}\ua}c^{\dag}_{k_{2}\da}+
    c^{\dag}_{k_{2}\ua}c^{\dag}_{k_{1}\da})-
    \frac{1}{2\sqrt{3}}( c^{\dag}_{k_{3}\ua}c^{\dag}_{k_{3}\da}-
     c^{\dag}_{k_{4}\ua}c^{\dag}_{k_{4}\da}+
      c^{\dag}_{k_{5}\ua}c^{\dag}_{k_{5}\da}-
       c^{\dag}_{k_{6}\ua}c^{\dag}_{k_{6}\da})\}|0\ket \\
    |\q^{^{1}\G_{1}}_{2}\ket=\frac{1}{2}(  
    c^{\dag}_{k_{3}\ua}c^{\dag}_{k_{5}\da}-
    c^{\dag}_{k_{4}\ua}c^{\dag}_{k_{6}\da}+
    c^{\dag}_{k_{5}\ua}c^{\dag}_{k_{3}\da}-
    c^{\dag}_{k_{6}\ua}c^{\dag}_{k_{4}\da})|0\ket 
    \end{array}
    \label{gamma1}
\end{equation}
\begin{equation}
    \begin{array}{l}
	 |\q^{^{1}\G_{2}}_{1}\ket=\{\frac{1}{\sqrt{3}}
	 ( c^{\dag}_{k_{1}\ua}c^{\dag}_{k_{1}\da}+
	  c^{\dag}_{k_{2}\ua}c^{\dag}_{k_{2}\da})+
	   \frac{1}{2\sqrt{3}}(  
    c^{\dag}_{k_{3}\ua}c^{\dag}_{k_{5}\da}+
    c^{\dag}_{k_{4}\ua}c^{\dag}_{k_{6}\da}+
    c^{\dag}_{k_{5}\ua}c^{\dag}_{k_{3}\da}+
    c^{\dag}_{k_{6}\ua}c^{\dag}_{k_{4}\da})\}|0\ket \\
    |\q^{^{1}\G_{2}}_{2}\ket=\frac{1}{2}(
    c^{\dag}_{k_{3}\ua}c^{\dag}_{k_{3}\da}+
     c^{\dag}_{k_{4}\ua}c^{\dag}_{k_{4}\da}+
      c^{\dag}_{k_{5}\ua}c^{\dag}_{k_{5}\da}+
       c^{\dag}_{k_{6}\ua}c^{\dag}_{k_{6}\da})|0\ket 
       \end{array}
        \label{gamma2}
\end{equation}
\begin{equation}
    \begin{array}{l}
|\q^{^{1}\S_{2}}_{1}\ket=\frac{1}{\sqrt{2}}(
c^{\dag}_{k_{1}\ua}c^{\dag}_{k_{1}\da}-
c^{\dag}_{k_{2}\ua}c^{\dag}_{k_{2}\da})|0\ket\\
|\q^{^{1}\S_{2}}_{2}\ket=\frac{1}{2}(
c^{\dag}_{k_{3}\ua}c^{\dag}_{k_{4}\da}+
c^{\dag}_{k_{4}\ua}c^{\dag}_{k_{3}\da}+
c^{\dag}_{k_{5}\ua}c^{\dag}_{k_{6}\da}+
c^{\dag}_{k_{6}\ua}c^{\dag}_{k_{5}\da})|0\ket\\
|\q^{^{1}\S_{2}}_{3}\ket=\frac{1}{2}(
c^{\dag}_{k_{3}\ua}c^{\dag}_{k_{6}\da}+
c^{\dag}_{k_{4}\ua}c^{\dag}_{k_{5}\da}+
c^{\dag}_{k_{5}\ua}c^{\dag}_{k_{4}\da}+
c^{\dag}_{k_{6}\ua}c^{\dag}_{k_{3}\da})|0\ket
\end{array}
 \label{sigma2}
\end{equation}
and finally
\begin{equation}
    \begin{array}{l}
|\q^{^{1}\W_{1}}_{1}\ket=\frac{1}{2}(
c^{\dag}_{k_{1}\ua}c^{\dag}_{k_{3}\da}+
c^{\dag}_{k_{3}\ua}c^{\dag}_{k_{1}\da}+
c^{\dag}_{k_{2}\ua}c^{\dag}_{k_{5}\da}+
c^{\dag}_{k_{5}\ua}c^{\dag}_{k_{2}\da})|0\ket\\
|\q^{^{1}\W_{1}}_{2}\ket=\frac{1}{2}(
c^{\dag}_{k_{1}\ua}c^{\dag}_{k_{4}\da}+
c^{\dag}_{k_{4}\ua}c^{\dag}_{k_{1}\da}-
c^{\dag}_{k_{2}\ua}c^{\dag}_{k_{6}\da}-
c^{\dag}_{k_{6}\ua}c^{\dag}_{k_{2}\da})|0\ket\\
|\q^{^{1}\W_{1}}_{3}\ket=\frac{1}{2}(
c^{\dag}_{k_{1}\ua}c^{\dag}_{k_{5}\da}+
c^{\dag}_{k_{5}\ua}c^{\dag}_{k_{1}\da}+
c^{\dag}_{k_{2}\ua}c^{\dag}_{k_{3}\da}+
c^{\dag}_{k_{3}\ua}c^{\dag}_{k_{2}\da})|0\ket\\
|\q^{^{1}\W_{1}}_{4}\ket=\frac{1}{2}(
c^{\dag}_{k_{1}\ua}c^{\dag}_{k_{6}\da}+
c^{\dag}_{k_{6}\ua}c^{\dag}_{k_{1}\da}-
c^{\dag}_{k_{2}\ua}c^{\dag}_{k_{4}\da}-
c^{\dag}_{k_{4}\ua}c^{\dag}_{k_{2}\da})|0\ket\\
|\q^{^{1}\W_{1}}_{5}\ket=\frac{1}{\sqrt{2}}(
c^{\dag}_{k_{3}\ua}c^{\dag}_{k_{3}\da}-
c^{\dag}_{k_{5}\ua}c^{\dag}_{k_{5}\da})|0\ket\\
|\q^{^{1}\W_{1}}_{6}\ket=\frac{1}{\sqrt{2}}(
c^{\dag}_{k_{4}\ua}c^{\dag}_{k_{4}\da}-
c^{\dag}_{k_{6}\ua}c^{\dag}_{k_{6}\da})|0\ket.
\end{array}
 \label{omega1}
\end{equation}
Other irreps yield $W=0$ triplet pairs. They are the three times degenerate irrep
\begin{equation}
    \begin{array}{l}
|\q^{^{3}\S_{3}}_{1}\ket=\frac{1}{\sqrt{2}}(
c^{\dag}_{k_{1}\ua}c^{\dag}_{k_{2}\da}-
c^{\dag}_{k_{2}\ua}c^{\dag}_{k_{1}\da})|0\ket\\
|\q^{^{3}\S_{3}}_{2}\ket=\frac{1}{2}(
c^{\dag}_{k_{3}\ua}c^{\dag}_{k_{4}\da}-
c^{\dag}_{k_{4}\ua}c^{\dag}_{k_{3}\da}+
c^{\dag}_{k_{5}\ua}c^{\dag}_{k_{6}\da}-
c^{\dag}_{k_{6}\ua}c^{\dag}_{k_{5}\da})|0\ket\\
|\q^{^{3}\S_{3}}_{3}\ket=\frac{1}{2}(
c^{\dag}_{k_{3}\ua}c^{\dag}_{k_{6}\da}-
c^{\dag}_{k_{4}\ua}c^{\dag}_{k_{5}\da}+
c^{\dag}_{k_{5}\ua}c^{\dag}_{k_{4}\da}-
c^{\dag}_{k_{6}\ua}c^{\dag}_{k_{3}\da})|0\ket
\end{array}
\end{equation}
and the two sixfold sets  
\begin{equation}
    \begin{array}{l}
|\q^{^{3}\W_{2}}_{1}\ket=\frac{1}{2}(
c^{\dag}_{k_{1}\ua}c^{\dag}_{k_{3}\da}-
c^{\dag}_{k_{3}\ua}c^{\dag}_{k_{1}\da}+
c^{\dag}_{k_{2}\ua}c^{\dag}_{k_{5}\da}-
c^{\dag}_{k_{5}\ua}c^{\dag}_{k_{2}\da})|0\ket\\
|\q^{^{3}\W_{2}}_{2}\ket=\frac{1}{2}(
c^{\dag}_{k_{1}\ua}c^{\dag}_{k_{4}\da}-
c^{\dag}_{k_{4}\ua}c^{\dag}_{k_{1}\da}-
c^{\dag}_{k_{2}\ua}c^{\dag}_{k_{6}\da}+
c^{\dag}_{k_{6}\ua}c^{\dag}_{k_{2}\da})|0\ket\\
|\q^{^{3}\W_{2}}_{3}\ket=\frac{1}{2}(
c^{\dag}_{k_{1}\ua}c^{\dag}_{k_{5}\da}-
c^{\dag}_{k_{5}\ua}c^{\dag}_{k_{1}\da}+
c^{\dag}_{k_{2}\ua}c^{\dag}_{k_{3}\da}-
c^{\dag}_{k_{3}\ua}c^{\dag}_{k_{2}\da})|0\ket\\
|\q^{^{3}\W_{2}}_{4}\ket=\frac{1}{2}(
c^{\dag}_{k_{1}\ua}c^{\dag}_{k_{6}\da}-
c^{\dag}_{k_{6}\ua}c^{\dag}_{k_{1}\da}-
c^{\dag}_{k_{2}\ua}c^{\dag}_{k_{4}\da}+
c^{\dag}_{k_{4}\ua}c^{\dag}_{k_{2}\da})|0\ket\\
|\q^{^{3}\W_{2}}_{5}\ket=\frac{1}{\sqrt{2}}(
c^{\dag}_{k_{3}\ua}c^{\dag}_{k_{5}\da}-
c^{\dag}_{k_{5}\ua}c^{\dag}_{k_{3}\da})|0\ket\\
|\q^{^{3}\W_{2}}_{6}\ket=\frac{1}{\sqrt{2}}(
c^{\dag}_{k_{4}\ua}c^{\dag}_{k_{6}\da}-
c^{\dag}_{k_{6}\ua}c^{\dag}_{k_{4}\da})|0\ket
\end{array}
\label{omega2}
\end{equation}
and
\begin{equation}
    \begin{array}{l}
|\q^{^{3}\W_{3}}_{1}\ket=\frac{1}{2}(
c^{\dag}_{k_{1}\ua}c^{\dag}_{k_{3}\da}-
c^{\dag}_{k_{3}\ua}c^{\dag}_{k_{1}\da}-
c^{\dag}_{k_{2}\ua}c^{\dag}_{k_{5}\da}+
c^{\dag}_{k_{5}\ua}c^{\dag}_{k_{2}\da})|0\ket\\
|\q^{^{3}\W_{3}}_{2}\ket=\frac{1}{2}(
c^{\dag}_{k_{1}\ua}c^{\dag}_{k_{4}\da}-
c^{\dag}_{k_{4}\ua}c^{\dag}_{k_{1}\da}+
c^{\dag}_{k_{2}\ua}c^{\dag}_{k_{6}\da}-
c^{\dag}_{k_{6}\ua}c^{\dag}_{k_{2}\da})|0\ket\\
|\q^{^{3}\W_{3}}_{3}\ket=\frac{1}{2}(
c^{\dag}_{k_{1}\ua}c^{\dag}_{k_{5}\da}-
c^{\dag}_{k_{5}\ua}c^{\dag}_{k_{1}\da}-
c^{\dag}_{k_{2}\ua}c^{\dag}_{k_{3}\da}+
c^{\dag}_{k_{3}\ua}c^{\dag}_{k_{2}\da})|0\ket\\
|\q^{^{3}\W_{3}}_{4}\ket=\frac{1}{2}(
c^{\dag}_{k_{1}\ua}c^{\dag}_{k_{6}\da}-
c^{\dag}_{k_{6}\ua}c^{\dag}_{k_{1}\da}+
c^{\dag}_{k_{2}\ua}c^{\dag}_{k_{4}\da}-
c^{\dag}_{k_{4}\ua}c^{\dag}_{k_{2}\da})|0\ket\\
|\q^{^{3}\W_{3}}_{5}\ket=\frac{1}{2}(
c^{\dag}_{k_{3}\ua}c^{\dag}_{k_{4}\da}-
c^{\dag}_{k_{4}\ua}c^{\dag}_{k_{3}\da}+
c^{\dag}_{k_{6}\ua}c^{\dag}_{k_{5}\da}-
c^{\dag}_{k_{5}\ua}c^{\dag}_{k_{6}\da}
)|0\ket\\
|\q^{^{3}\W_{3}}_{6}\ket=\frac{1}{2}(
c^{\dag}_{k_{3}\ua}c^{\dag}_{k_{6}\da}-
c^{\dag}_{k_{6}\ua}c^{\dag}_{k_{3}\da}+
c^{\dag}_{k_{4}\ua}c^{\dag}_{k_{5}\da}-
c^{\dag}_{k_{5}\ua}c^{\dag}_{k_{4}\da}
)|0\ket.
\end{array}
\label{omega3}
\end{equation}
The above {\em a priori} argument hardly applies to the symmetries of 
$W=0$ quartets, because $\Omega_{4}^{4}$ contains almost every 
symmetry and we do not know any $W=0$-like {\it Theorem} for quartets. 
However, we can still build the projection operators by Mathematica; 
we can project the 225 quartets on the irreps of ${\cal G}$  
and carry on the analysis in an efficient, if not elegant, way.
We found that the singlet $W=0$ quartets are 13 as many as the 
singlet $W=0$ pairs, and belong to the same irreps
$\G_{1},\;\G_{2},\;\S_{2},\;\W_{1}$. Therefore, these are the 
possible symmetries of the first-order ground states with  14 holes. 
Exact diagonalization results\cite{fop} show that for $U/t<3$ and 
$16-2=14$ holes the ground state is sixfold degenerate, with a 
doublet of states with momentum $(\p,\p)$ and a quartet with 
momentum $(\pm\p/2,\pm\p/2)$. In  view of Eqs.(\ref{omega1}), the computed ground 
state corresponds to an $\W_{1}$ {\it electron} pair over the half 
filled system. For $U/t>3$ and the same number of 
holes a level crossing takes place: 
the ground state is threefold degenerate and contains a 
state with momentum $(0,0)$ and a doublet with momentum $(\p,0)$ 
and $(0,\p)$. In  view of Eqs.(\ref{sigma2}),
the computed ground 
state must be assigned to a $\S_{2}$ {\it electron} pair over the half 
filled system. In both cases, the symmetry of the 
ground state corresponds to a $W=0$ pair. 

\section {Local picture of the $W=0$ Pairs and Quartets}
\label{quartetti}

The {\em antiferromagnetic property} of the local 
basis of any  site $r$  readily prompts  $W=0$ pairs; 
alternatively, we can  transform the pairs and quartets of 
well-defined symmetry (Section \ref{12h}) using the  local basis.  Both methods are useful. 
The vacuum at half filling is a  $W=0$  state with 6 holes.
The  Quartets  are then obtained from the antiferromagnetic ground 
state at half filling by removing a  $W=0$ pair.
 
\subsection{W=0 Pairs}
\label{W=0 Pairs}

Chosen a site $r$, let $|\phi^{(r)}_{\alpha}\ket$ be a normalized linear 
combination of the states $|\phi^{(r)}_{1}\ket,|\phi^{(r)}_{2}\ket,
|\phi^{(r)}_{3}\ket$ of ${\cal S}_{a}$ and
$\phi^{(r)}_{\beta}$ be a normalized linear 
combination of 
$|\phi^{(r)}_{4}\ket,|\phi^{(r)}_{5}\ket,|\phi^{(r)}_{6}\ket$ 
of ${\cal S}_{b}$. Then, 
\begin{equation}
|d^{(r)}_{\alpha,\beta}\ket= 
|\phi^{(r)}_{\alpha\ua}\phi^{(r)}_{\beta\da}\ket
\label{coppia}
\end{equation}
is a two-body state free of double occupation on every site and
$\frac{|\phi^{(r)}_{\alpha\ua}\phi^{(r)}_{\beta\da}\ket+
|\phi^{(r)}_{\beta\ua}\phi^{(r)}_{\alpha\da}\ket}{\sqrt{2}}$ is a 
$W=0$ pair. Since for each spin one has 3 degrees of freedom, one can build 
9 independent pairs in this way; they are bases for the $\Omega_{1}$ and 
$\Sigma_{2}$ $W=0$ pairs of the previous Section. 
The only alternative method for obtaining $W=0$ pairs is that of forming  
$|\phi^{(r)}_{\alpha}\ket$ and  $|\phi^{(r)}_{\beta}\ket$
as linear combinations of states of the same subset (both from 
${\cal S}_{a}$ or both from  ${\cal S}_{b}$). This can be accomplished in 
such a way that $|\phi^{(r)}_{1}\ket$ never appears for both spins; 
actually, the bases of $\Gamma_{1}$ and $\Gamma_{2}$ are obtained in 
this way. If we use ${\cal S}_{a}$  for both spin directions then 
after a lattice step in any direction the pair is formed exclusively 
with states of ${\cal S}_{b}$ and its occupation vanishes;  such pairs 
{\em live} on a sublattice.

We rewrite the pairs and quartets of Section 
\ref{12h} using the 
local basis of any  site $r$. By Eq.(\ref{inverso}),
\begin{equation}
|k_{i}\ua k_{j}\da\ket =\sum_{m,n=1}^{6}e^{-i(k_{i}+k_{j})r}O_{ni}O_{mj}
|\f_{n\ua}^{(r)}\f_{m\da}^{(r)}\ket.
\label{percoppie}
\end{equation}
For instance, the operator that annihilates the 6-$th$ component of 
the $\Omega_{1}$ $W=0$ pair in Eq.(\ref{omega1}) becomes
\begin{equation}
\Q_{6}^{^{1}\Omega_{1}}=-\frac{1}{\sqrt{6}}(c_{1\ua}c_{6\da}+c_{6\ua}c_{1\da})+
\frac{1}{2}(c_{6\ua}c_{3\da}+c_{3\ua}c_{6\da})+\frac{1}{2\sqrt{3}}
(c_{2\ua}c_{6\da}+c_{6\ua}c_{2\da}),
\label{sesta}
\end{equation} 
where $c^{(r)^{\dag}}_{i\s}$ and $c^{(r)}_{i\s}$ are hole creation and 
annihilation operators in the local states $\f^{(r)}_{i\s}$; in 
Eq.(\ref{sesta}) the site is not specified since whatever it is 
$\Q_{6}^{^{1}\Omega_{1}}$ does not change. 
The local representation of symmetry-adapted pairs is  of 
interest because different irreps are well characterized by their 
local behaviour.

\subsection{W=0 Quartets}
\label{W=0 Quartets}

With 14 holes, in the $U \rightarrow 0$ limit, there are four in ${\cal 
S}_{hf}$; the first-order ground states correspond to $W=0$ quartets.
By removing in all possible ways two holes of opposite spin from the 6-body 
$W=0$ determinant $|d^{(r)}[1,2,3]\ket_{\s}$ of 
equation (\ref{antidiag}) one produces nine 4-body determinants.  They 
are free of double occupation on site $r$ because  $|\phi^{(r)}_{1}\ket$  
cannot appear for both spins; this property holds on all sites 
because of the special form of the translation matrices 
(\ref{translationx}) and (\ref{translationy}).  It follows that these 
are $W=0$  
states and  in first-order perturbation theory they belong to the ground 
state multiplet. It is clear that these 9 first-order ground states are 
in one-to-one correspondence to the pairs (\ref{coppia}), and for 
instance 
$|\phi^{(r)}_{2\da}\phi^{(r)}_{3\da}\phi^{(r)}_{5\ua}
\phi^{(r)}_{6\ua}\ket$ corresponds to the pair
$|\phi^{(r)}_{1\da}\phi^{(r)}_{4\ua}\ket$;
since $|\F_{AF}^{S=0}\ket$ transforms as the totalsymmetric 
one-dimensional irrep $A_{1}$ under the 
$C_{4v}$ operations referred to the center of a plaquette, the symmetries of 
the 9 quartets and the 9 pairs are also the same , namely, they are 
$\Omega_{1}$ and $\Sigma_{2}$ of ${\cal G}$. The total momentun 
labels are also the same. There is actually a complete correspondence 
between $W=0$ pairs and 
quartets; the quartets are also 13 and those that were not obtained 
above belong to $\Gamma_{1}$ and $\Gamma_{2}$. These cannot arise in 
the same way because one can show that they are not obtained by 
removing two holes from $|d^{(r)}[1,2,3]\ket_{\s}$. This means that 
$\Gamma_{1}$ and $\Gamma_{2}$ are not to be interpreted as pairing 
states.

\section{Pairing mechanism}
\label{mechanism}
 
We consider the ground state of the $4\times 4$ model with 14 holes; aside 
from the 10 holes in the inner $A_{1}$ and $\Lambda_{1}$ shells (see 
Table IV) the  outer $\W_{4}$ shell contains 4 holes in a $W=0$ 
quartet. We are in position to show that the principles of
Section \ref{smallcluster} produce a diagnosis of pairing 
between two {\em electrons} added to the  antiferromagnetic 16-holes ground 
state (half filling). 

We recall from Section \ref{12h} that, by  comparing with exact 
diagonalization results\cite{fop}, the ground state is assigned 
to  $\W_{1}$, Eq.(\ref{omega1}), at weak coupling and to $\S_{2}$,  
Eq.(\ref{sigma2}), at a stronger coupling.
 In analogy with the points $a$)-$c$) of Section \ref{smallcluster}, we must 
preliminarily verify that  symmetry does not forbid  obtaining these 
symmetries by creating  $W=0$ electron pairs, 
Eqs.(\ref{gamma1}-\ref{omega3}),  
from the antiferromagnetic state of Eq.(\ref{af}). This is the same as annihilating 
hole pairs. Since the state $|\F_{AF}^{S=0}\ket$ of Eq.(\ref{af}) is a total symmetric 
singlet with vanishing total momentum, the labels of the quartets will be the same of 
the annihilated hole pairs. This operation can be done by hand, or with the help 
of Mathematica, and the answer to the preliminary question is 
adfirmative for $\W_{1}$ and $\S_{2}$, but not for all pairs. 
We obtain 24 $W=0$ quartets 
out of the 28 states in Eqs.(\ref{gamma1}-\ref{omega3}), since 
the annihilation of $\G_{1}$ and 
$\G_{2}$ $W=0$ pairs gives identically zero.

We need to modify the  canonical transformation  to  deal with the  
 $4\times4$ cluster with periodic boundary conditions near half filling; 
 the previous form is not adequate because the
 ${\cal S}_{hf}$ shell is only partially occupied  
and so in Eqs.(\ref{statim},\ref{statia}) we have to use the antiferromagnetic 
state $|\F_{AF}^{S=0}\ket$ of Eq.(\ref{af}). 
In small clusters like the $4\times 4$ one the one-body states are 
widely separated and the intra-shell interaction is much more 
important than the inter-shell one; therefore, we consider only the $m$ 
states made removing two holes in ${\cal S}_{hf}$ from $|\F_{AF}^{S=0}\ket$, 
neglecting the high-lying unoccupied orbitals\cite{cbs3}. 

We now come to item $d$) of Section \ref{smallcluster}, the identification of 
the two states which are obtained by spin-flip from each other and 
are coupled by the interaction in second order. 
The explicit  form of the symmetry adapted pairs of Section \ref{12h} 
shows that $|\q^{^{1}\S_{2}}_{1}\ket$ and  $|\q^{^{3}\S_{3}}_{1}\ket$ involve 
only $k_{1}$ and $k_{2}$. This suggests that by taking linear 
combinations we can obtain  a two-body state and its spin-flipped 
image. Specifically, using the electron creation operators 
\begin{equation}
     \a^{[\pm]}_{\s} \equiv \frac{1}{\sqrt{2}}(c_{k_{1}\s}\pm 
     c_{k_{2}\s})
\label{12pm}
\end{equation}
we build the following two-electron $W=0$ determinants:
\begin{equation}
    |w_{1}\ket= \a^{[+]}_{\ua}\a^{[-]}_{\da}|0\ket_{e} \quad,\quad 
    |w_{2}\ket= \a^{[-]}_{\ua}\a^{[+]}_{\da}|0\ket_{e}.
   \label{ww}
\end{equation}
where the state $|0\ket_{e}$ is the electron-vacuum state: 
$c^{\dag}_{k\s}|0\ket_{e}=0,\;\forall k,\s$. 
With Eq.(\ref{ww}) we can build the following $m$ states:
\begin{equation}
    |m_{1}\ket= \a^{[+]}_{\ua}\a^{[-]}_{\da}|\F_{AF}^{S=0}\ket \quad,\quad 
    |m_{2}\ket= \a^{[-]}_{\ua}\a^{[+]}_{\da}|\F_{AF}^{S=0}\ket.
\label{mnuovi1}
\end{equation} 
Since $|m_{1}\ket$ and $|m_{2}\ket$ have projection only on irreps 
$\S_{2}$ and $\S_{3}$,
the states (\ref{mnuovi1}) can be mixed by the operator 
$W_{eff}$ only between themselves, and so they are the only states 
involved in the canonical transformation. The eigenvalue equation 
(\ref{can1}) reduces to a $2\times2$ problem; the  $W_{eff}$ matrix 
to diagonalize is: 
\begin{equation}
\left[\begin{array}{rr}
0\quad\quad\; & \bra m_{1}|W_{eff}|m_{2}\ket \\
\bra m_{2}|W_{eff}|m_{1}\ket & 0\quad\quad\;  \end{array}\right] .
\label{weffmat1}
\end{equation}

The eigenvalues are $\pm \bra m_{1}|W_{eff}|m_{2}\ket $ and the eigenvectors are 
$\frac{1}{\sqrt{2}}(1,\pm 1)$. Expanding these 
eigenvectors in the base (\ref{mnuovi1}), we find:
\begin{equation}
     |\Psi_{1}^{^{1}\S_{2}}\ket 
     =\frac{1}{\sqrt{2}}(|m_{1}\ket+|m_{2}\ket) \eq 
     |\q^{^{1}\S_{2}}_{1}\ket_{e} \otimes |\F_{AF}^{S=0}\ket,\quad 
     |\Psi_{1}^{^{3}\S_{3}}\ket 
     =\frac{1}{\sqrt{2}}(|m_{1}\ket-|m_{2}\ket) \eq |\q^{^{3}\S_{3}}_{1}\ket_{e}
     \otimes|\F_{AF}^{S=0}\ket,
 \label{st1}
\end{equation}
where the notation $|\q^{\h}_{i}\ket_{e}\otimes|\F_{AF}^{S=0}\ket$
stands for creating the electron-pair $|\q^{\h}_{i}\ket$ 
over the antiferromagnetic state $|\F_{AF}^{S=0}\ket$ (filled shells are 
understood) to get a 14 hole state.
We use this example to stress that  pairing by 
the present mechanism is possible only if the Optimal Group  ${\cal G}$ possesses 
{\em twin} singlet and triplet representations with the same number of 
components, built by the same orbitals.

In a similar way, $|\q^{^{1}\W_{1}}_{6}\ket$ and  $|\q^{^{3}\W_{2}}_{6}\ket$ involve 
only $k_{4}$ and $k_{6}$; this suggests introducing the electron 
creation operators
\begin{equation}
     \b^{[\pm]}_{\s}
     \equiv \frac{1}{\sqrt{2}}(c_{k_{4}\s}\pm c_{k_{6}\s});
\label{46pm}
\end{equation}
the  two-electron $W=0$ determinants
\begin{equation}
    |w'_{1}\ket= \b^{[+]}_{\ua}\b^{[-]}_{\da}|0\ket_{e} \quad,\quad 
    |w'_{2}\ket= \b^{[-]}_{\ua}\b^{[+]}_{\da}|0\ket_{e}
   \label{www}
\end{equation}
are obtained from each other by spin-flip. Incidentally, the latter property is 
preserved if one switches to the local picture by
the transformation (\ref{inverso}): introducing a new orbital 
\begin{equation}
c_{\Omega_{6}\s}= 
 \frac{-c_{1\s}}{\sqrt{3}}+
 \frac{c_{5\s}}{\sqrt{2}}+
 \frac{c_{6\s}}{\sqrt{6}}
 \label{1body}
\end{equation} 
one finds
\begin{equation}
 |w'_{1}\ket =c_{\Omega_{6}\ua}c_{6\da}|0\ket_{e},
 \label{mvecchio1}
\end{equation} 
while
\begin{equation}
 |w'_{2}\ket =c_{6\ua}c_{\W_{6}\da}|0\ket_{e}.
 \label{mvecchio2}
\end{equation} 
Thus, we have identified the  $m$ states
\begin{equation}
    |m'_{1}\ket= \b^{[+]}_{\ua}\b^{[-]}_{\da}|\F_{AF}^{S=0}\ket \quad,\quad 
    |m'_{2}\ket= \b^{[-]}_{\ua}\b^{[+]}_{\da}|\F_{AF}^{S=0}\ket.\label{mnuovi2}
\end{equation}
Since $|m'_{1}\ket$ and $|m'_{2}\ket$  have projection only on irreps  
$\W_{1}$ and $\W_{2}$, the states (\ref{mnuovi2}) can be mixed by the operator 
$W_{eff}$ only between themselves, and we find another $2 \times 2$ 
matrix
\begin{equation}
 \left[\begin{array}{rr}
0\quad\quad\; & \bra m'_{1}|W_{eff}|m'_{2}\ket \\
\bra m'_{2}|W_{eff}|m'_{1}\ket & 0\quad\quad\;  \end{array}\right]
\label{weffmat2}
\end{equation}
with  eigenvalues $\pm \bra m'_{1}|W_{eff}|m'_{2}\ket$ and eigenvectors 
$\frac{1}{\sqrt{2}}(1,\pm 1)$. Therefore,
\begin{equation}
 |\Psi_{2}^{^{1}\W_{1}}\ket 
     =\frac{1}{\sqrt{2}}(|m'_{1}\ket+|m'_{2}\ket) \eq 
     |\q^{^{1}\W_{1}}_{6}\ket_{e}\otimes|\F_{AF}^{S=0}\ket,\quad 
     |\Psi_{6}^{^{3}\W_{2}}\ket 
     =\frac{1}{\sqrt{2}}(|m'_{1}\ket-|m'_{2}\ket) \eq 
     |\q^{^{3}\W_{2}}_{6}\ket_{e} \otimes |\F_{AF}^{S=0}\ket.
 \label{st2}
\end{equation}
The numerical values of the eigenvalues for $U=-t=1\, eV$ 
are negative in both cases,
\begin{equation}
\bra m_{1}|W_{eff}|m_{2}\ket=-60.7\,meV \quad\quad 	
\bra m'_{1}|W_{eff}|m'_{2}\ket=-61.9\,meV
\end{equation}
which means that in both the cases the Cooper-like equation (\ref{can1}) gives 
singlet ground states; the triplets have an opposite correction to the energy.
The binding energy is larger for the  
$|\q^{^{1}\W_{1}}\ket$ singlet, which is the true ground state (the 
result cannot depend on the component of the irrep). 
Moreover since $|\F_{AF}^{S=0}\ket$ is a totalsymmetric 
singlet with vanishing total momentum, we can also predict momentum, spin and 
symmetry of the ground state.
Our approach has enough predictive power to 
yield the symmetry, wavevector  and degeneracy of the ground state.
\section{Conclusions}
\label{conc}
Recently, we have proposed a new approach\cite{cbs1},\cite{cbs2}
to the question of the existence and nature of Cooper-like pairing
from repulsive interactions; this is based on a configuration mixing
realized by a canonical 
transformation; the {\em vacuum} is the ground state of the system 
at $U=0^{+}$ and the Cooper-like bound states arise from $W=0$ pairs.
Thus our approach is actually { \underline not} a perturbation theory 
but its application is easier in weak coupling situations, when one 
can limit the configuration mixing to a few particle-hole excitations.
The pairing mechanism involved  is a form of 
spin-flip exchange diagram which is enhanced by  the $C_{4v}$ symmetry and 
our canonical 
transformation works independenly of the size of the system. We have 
tested the results with exact diagonalization data for open-boundary 
condition clusters \cite{cibal}\cite{cbs3} in the three-band Hubbard model. 
We obtained good agreement with the numerical data already by the 
simplest approximation within our scheme, i.e. by truncating the 
configuration mixing at the level of electron-hole pairs exchange.

In this paper we test our $W=0$ pairing mechanism within the one-band Hubbard 
model with periodic boundary condition using exact  numerical data on the 
$4\times 4$ square lattice\cite{fop}. In particular, several workers have 
recognised that those data could be qualitatively understood by a weak 
coupling analysis\cite{fridman}\cite{galan}; however there was no 
clear-cut conclusion about the existence of pairing and the symmetry 
analysis of the system was not complete.

We point out the criteria that allow  one to unambiguously diagnose pairing 
of two holes added to the system. At the heart 
of the effective interaction there is the diagram of Figure 1. To 
characterize the symmetry of the  ground state with 2 added holes we 
obtained the analytic ground state at  
half filling for $U=0^{+}$ by a new approach\cite{antif}. Moreover, we 
need the symmetry of the $W=0$ pairs in the system at hand. To this 
end we prove a general result that we call $W=0$ Theorem. The full 
exploitation of this theorem requires the knowledge of the Optimal 
Group, that is, a  symmetry group which is so big that no 
accidental degeneracies arise. We have obtained this Group here for 
the first time. An adequate knowledge of the symmetry of the system 
allowed us to develop in detail the canonical trasformation on the 
$4\times 4$ cluster analytically; thus we obtained good agreement with the 
data and a clear diagnosis of pairing.  This result lends further 
support to the general approach\cite{cbs1},\cite{cbs2} which predicts 
pairing in the $N \times N$ cluster for any $N$; moreover, we stress 
that we have already shown how the the $W=0$ pairing grants the 
superconducting flux quantization \cite{cbs3}.

\section{Acknowledgements}

This work was supported by the Istituto Nazionale di Fisica della 
Materia.

\appendix

\section{Contributions to the $W$ matrix from Filled Shells }
\label{equivalence}

The $ N^{2}-$body determinantal wave functions with  
$S_{z}=0$ that one can build using the orbitals with  $\e(k)<0$ and 
half of those with $\e(k)=0$ are a set of  $\left(\begin{array}{c} 2N-2 \\ N-1 \end{array}\right)^{2}$
elements. Each represents one of the  degenerate unperturbed ($U=0$) ground state 
configurations at half filling. First-order perturbation theory requires the diagonalization of 
the $W$ matrix over such a basis.

The diagonal elements of the $W$ matrix are just expectation values over 
determinants $|k_{\a}\ua k_{\b} \da \ldots \ket$.
Such an expectation value is a  sum over all the possible pairs of the 
bielectronic elements of $W$ like 
\begin{equation}
W(\a \b,\a \b)=\sum_{r}U\bra k_{\a}|n_{r}|k_{\a}\ket\bra 
k_{\b}|n_{r}|k_{\b}\ket =
\sum_{r}U\frac{1}{N^{2}}e^{i(k_{\a}-k_{\a})r}\frac{1}{N^{2}}e^{i(k_{\b}-k_{\b})r}=
\frac{U}{N^{4}}N^{2}=\frac{U}{N^{2}};
\end{equation}
the result is independent of   $k_{\a}$ and 
$k_{\b}$. Since in any determinant of the set $N^{2}/2$ plane wave states are occupied for each 
spin, there are 
$N^{4}/4$ pairs, and the diagonal elements are all equal to  $UN^{2}/4$. 
Thus, the diagonal elements shift all the eigenvalues by this fixed 
amount.

The off-diagonal elements of the $W$ matrix between determinants that 
differ by three or more spin-orbitals vanish because $W$ is a two-body 
operator. The off-diagonal elements  between determinants that 
differ by one spin-orbital are sum of contributions like 
$W(\a \b,\gamma \b)=\sum_{r}U\bra k_{\a}|n_{r}|k_{\g}\ket\bra 
k_{\b}|n_{r}|k_{\b}\ket$ that vanish because of the orthogonality of the 
plane-wave orbitals. One is left with  the off-diagonal elements  between determinants that 
differ by two spin-orbitals, which coincide with the corresponding bielectronic 
elements  $W(\a \beta,\gamma \delta)=\sum_{r}U\bra k_{\a}|n_{r}|k_{\g}\ket\bra 
k_{\b}|n_{r}|k_{\d}\ket$.  This is just  the matrix of $W$ over 
the truncated Hilbert space ${\cal H}$ spanned by the states of the 
holes in the half filled shell, ignoring the filled ones. We stress 
that  there are $N-1$ 
holes of each spin in ${\cal S}_{hf}$, thus ${\cal H}$ is much 
smaller than the full Hilbert space of the Hubbard Hamiltonian; 
however, since the  number of holes grows linearly with $N$, the 
problem is still far from trivial.


\begin{references}
\bibitem{appendix} $C_{4v}$ is the symmetry group of a square. 
It is a finite group of 
order 8 and it contains 4 one dimensional irreps, 
$A_{1},\;A_{2},\;B_{1},\;B_{2}$, and 1 two-dimensional one called $E$. 
The table of characters is
\begin{center}
    \vspace*{0.5 cm}
\begin{tabular}{|c|c|c|c|c|c|}
\hline 
$C_{4v}$ & $\mathbf{1}$ & $C_{2}$ & $C^{(+)}_{4},\,C^{(-)}_{4}$ & 
$\s_{x},\,\s_{y}$ & 
$\s'_{+},\,\s'_{-}$ \\
\hline 
$A_{1}$ & 1 & 1 & 1 & 1 & 1 \\
\hline 
$A_{2}$ & 1 & 1 & 1 & -1 & -1 \\
\hline 
$B_{1}$ & 1 & 1 & -1 & 1 & -1 \\
\hline 
$B_{2}$ & 1 & 1 & -1 & -1 & 1 \\
\hline 
$E$ & 2 & -2 & 0 & 0 & 0 \\
\hline 
\end{tabular}
\vspace*{0.5 cm}
\end{center}
\bibitem{he} Equivalently, we shall find the exact 6-body ground state of the  
effective hamiltonian 
\begin{equation}
H_{eff}=\frac{U}{4^{4}}\sum_{k_{1},k_{2},k_{3},k_{4}\in {\cal S}_{hf}}
\d(k_{1}+k_{2}-k_{3}-k_{4})
c^{\dag}_{k_{1}\ua}c^{\dag}_{k_{2}\da}c_{k_{3}\da}c_{k_{4}\ua}
\label{heff}
\end{equation}
where $\d(G)=1$ if $G$ is a reciprocal lattice vector and zero 
otherwise.
\end{references}
\end{document}